%% file: main.tex
\documentclass[conference]{IEEEtran}

\usepackage{graphicx}
\graphicspath{{../pdf/}{../jpeg/}}
\DeclareGraphicsExtensions{.pdf,.jpeg,.png}

\usepackage{caption}
\newcommand{\figref}[1]{Figure~\ref{#1}}

\usepackage{amsmath} 
\usepackage{algorithmic} 
\usepackage{array} 
\usepackage{booktabs} 
\usepackage{multirow} 
\usepackage{url} 
\usepackage{geometry} 
\usepackage{xcolor} 
\definecolor{mycolor}{RGB}{255,0,0}

\usepackage{hyperref}

\usepackage{mdwmath}
\usepackage{mdwtab}
\usepackage{eqparbox}
\usepackage[nocompress]{cite} 
\usepackage{academicons} 
\usepackage{tikz} 
\usepackage{tabularx} 
\usepackage{amssymb} 
\usepackage{lettrine} 

\ifdefined\labelindent
\else
  \newlength{\labelindent}
\fi

\begin{document}

\title{

\vspace{-1cm} 
    \begin{tikzpicture}[remember picture, overlay]
        \node[anchor=north, yshift=-0.5cm] at (current page.north) {\fbox{\parbox{\textwidth}{\centering\small {\color{red}This work has been submitted to the IEEE Conference for possible publication. Copyright may be transferred without notice, after which this version may no longer be accessible.}}}};
    \end{tikzpicture}
    \vspace{0cm} 

RANFusion: A Comprehensive Tool for Simulating Handover In Next-G RAN}

\author{
    \IEEEauthorblockN{
        Seyed Bagher Hashemi Natanzi\IEEEauthorrefmark{1},
        Bo Tang\IEEEauthorrefmark{1}
    }
    \IEEEauthorblockA{
        \IEEEauthorrefmark{1}Electrical and Computer Engineering, Worcester Polytechnic Institute, Massachusetts, USA\\
        Email: \{snatanzi, btang1\}@wpi.edu
    }

    }
\definecolor{lime}{HTML}{A6CE39}
\DeclareRobustCommand{\orcidicon}{%
	\begin{tikzpicture}
	\draw[lime, fill=lime] (0,0) 
	circle [radius=0.16] 
	node[white] {\href{https://orcid.org/0000-0000-0000-0000}{\tiny ID}};
	\draw[white, fill=white] (-0.0625,0.095) 
	circle [radius=0.007];
	\end{tikzpicture}
	\hspace{-2mm}
}

\newcommand{\orcidauthor}[2]{%
	\href{https://orcid.org/#2}{#1\,\orcidicon}
}

\newcommand{\corauthor}[2]{%
	\href{https://orcid.org/#2}{#1* \,\orcidicon}
}
\maketitle
\input{include/abstract}
\input{include/Introduction}
\input{include/Background_and_Related_Work}
\input{include/Handover_workflow}
\input{include/RANFusion_System_Design_and_Architecture}
\input{include/Experiment_Result}

\vspace{-2pt}
\vspace{-2pt}
\input{include/Future_research_directions_and_potential_enhancements_to_RANFusion}



\section*{\textcolor{black}{Acknowledgment}}
\noindent
This work was supported in part by NSF awards CNS-2120442 and IIS-2325863. Supplement references with recent works in AI-driven open RAN research.


\bibliographystyle{IEEEtran}

\bibliography{./bib/main.bib}

\end{document}

%% file: include/abstract.tex
\begin{abstract}
The rapid advancement of 5G networks and the upcoming transition to 6G necessitate the use of the Open Radio Access Network (O-RAN) architecture to enable greater flexibility, interoperability, and innovation. This shift towards 6G and O-RAN requires the development of advanced simulation tools for testing, analyzing, and optimizing Radio Access Network (RAN) operations. This need becomes critical due to the complex dynamics of mobility management inherent in the 6G vision and next-generation networks. These networks anticipate advanced handover methods for mobile users, UAVs, IoT devices, and beyond. Addressing this gap, this paper introduces RANFusion: a robust RAN simulator specifically created to explore a variety of handover scenarios and to test and balance resources between users. This tool enables precise simulations for refining handover strategies within RAN and O-RAN environments, thereby ensuring optimal performance and reliability in these advanced network infrastructures.

\end{abstract}

\IEEEoverridecommandlockouts
\begin{keywords}
5G, 6G, Open RAN, O-RAN, RAN, Handover
\end{keywords}

\IEEEpeerreviewmaketitle

%% file: include/Introduction.tex
\section{Introduction}
\begin{figure*}[!t] 
\centering
\includegraphics[width=1\textwidth]{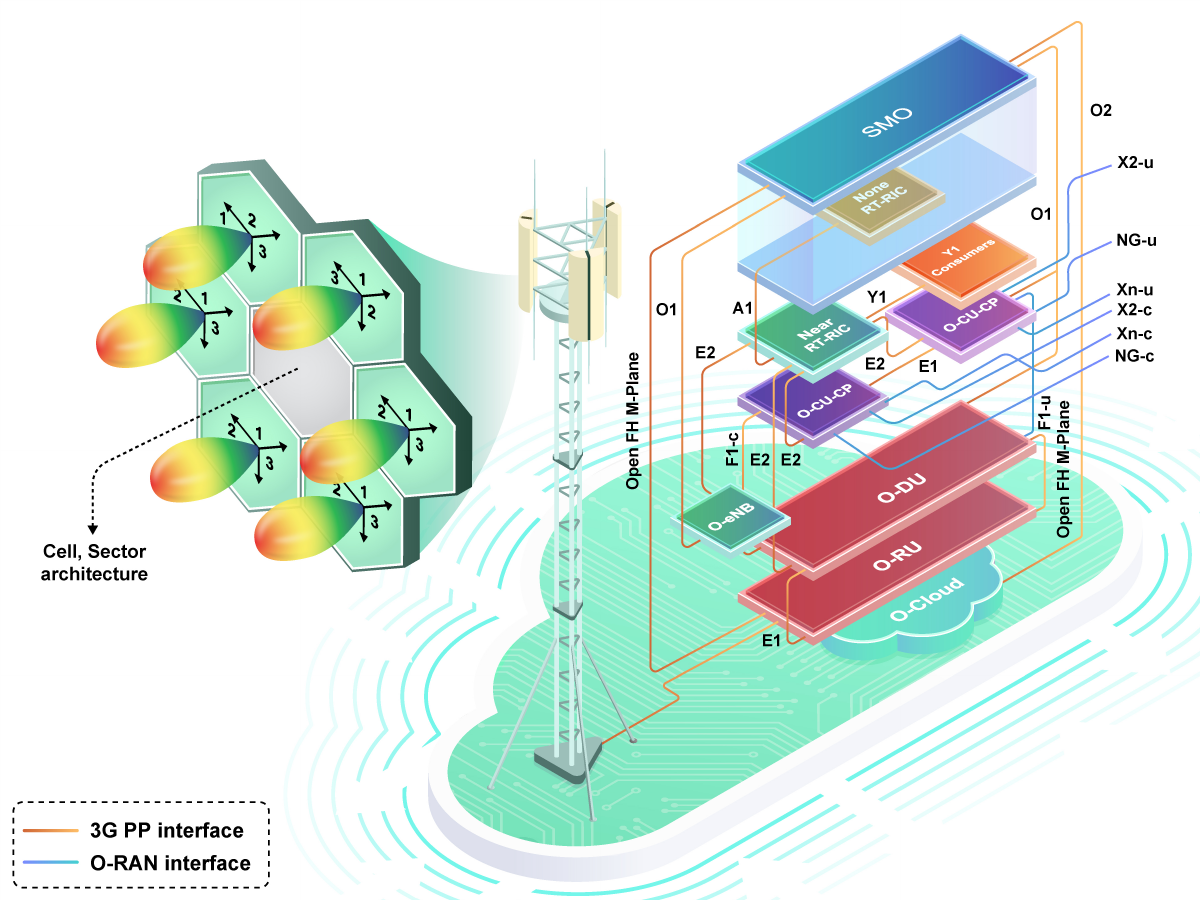}
\caption{Cell Sectorization Architecture}
\label{Figure:ORAN-arc}
\end{figure*}


\lettrine[nindent=1.9em,lines=1]{T}{\textnormal{he rapid}} evolution of mobile networks from 5G to 6G has initiated a paradigm shift in RAN design and operation. Adopting the O-RAN architecture is crucial for enhancing flexibility, interoperability, and innovation in both 5G and 6G networks. \cite{lin2023fundamentals}. The RAN Intelligent Controller (RIC) consists of near-real-time components (near-RT-RIC) and non-real-time components (non-RT-RIC). Developers can design applications (xApps) that directly controls the RAN element and integrate Artificial Intelligence (AI) and Machine Learning (ML) with the network. This distinction is pivotal; near-RT RIC focuses on orchestrating operations within seconds (typically 10ms to 1s) for dynamic policy enforcement and optimization, whereas non-RT RIC operates on a timescale of minutes to hours, addressing less time-sensitive control and management functions. For next-generation cellular networks (6G and beyond) and their use cases, which will incorporate non-terrestrial networks, such as satellites and Unmanned Aerial Vehicle (UAV) base stations  \cite{10279738}, alongside terrestrial networks \cite{panitsas2024predictive}, low-latency and high-reliability services are needed \cite{DAO2024110370}. This will increase the number of handovers due to smaller coverage areas and higher signal attenuation.
 Meanwhile, the transition to O-RAN infrastructure introduces new challenges, particularly in mobility management and handover processes. These processes are essential for ensuring smooth connections and the efficient use of resources.

\vspace{-1pt} To address handover challenges, simulation becomes crucial. However, despite the availability of many RAN simulators, existing tools often face scalability issues, especially in scenarios involving more than two user equipment (UE) or multiple gNodeBs. These issues are particularly problematic in situations where handovers are expected to occur seamlessly without signal weakness, changing the interference levels, or under specific conditions. Problems such as ping-pong effects, failed handovers, and late handovers necessitate robust simulation tools for testing and enhancing algorithms\cite{Haghrah2023}.

\vspace{-1pt} This paper presents RANFusion, a solution designed to improve handover in RAN and O-RAN simulations, especially for difficult handover situations where the limited E2 interface support and scalability limits of UEs and gNodeBs make things harder. RANFusion follows 3GPP standards and improves the testing framework by providing comprehensive support for advanced load-balancing and effective handover strategy assessments, crucial for xApp and rApp use cases. It supports an unlimited number of UEs, gNodeBs, cells, and sectors, and isolates the network to enable precise, customizable testing through JSON-driven configuration files. Significant improvements in the implementation phase include integrating a semi-functional E2 interface (E2-like) and exposing an API to the xApp, which enhances xApp scenario testing during handovers. \figref{Figure:ORAN-arc} shows the latest O-RAN architecture proposed by the O-RAN Alliance, and RANFusion's cell and sector designs.

\vspace{1pt}The paper is organized as follows: Section II reviews the related literature, Section III elaborates on the handover process, Section IV details the design and architecture of RANFusion, Section V showcases the results, and Section VI provides the concluding remarks. RANFusion's source code and documentation are available on GitHub\footnote{RANFusion source code and documentation: \url{https://github.com/openaicellular/oaic-t/tree/master/RANFusion}}.

%% file: include/Background_and_Related_Work.tex
\section{Related Work}

The RAN simulator have been presented in several works, as summarized in Table \ref{tab:RAN Simulators Comparison}. Our focus is on open-source academic software. Open Air Interface (OAI) and srsRAN, which support 4G/5G stacks, are limited due to their lack of full E2 support, a critical component for effective O-RAN handovers. E2E simulators and OMNeT++-based models, although facing limitations in supporting dynamic and scalable network scenarios under O-RAN standards, still offer enhancements \cite{PATRICIELLO2019101933}. Nardini et al. introduced an OMNeT++-based model library to simulate 5G networks, and the GTEC 5G link-level simulator presents features but is limited in handover scenarios \cite{9211504,7801585}. The my5G tester and 5G-air provide a simulation module for 5GNR but do not support O-RAN \cite{SILVEIRA2022109301,MARTIRADONNA2020107314}. Additionally, \cite{boeira2024calibrated} presents the calibrated Simu5G with limited scenarios.
\begin{table}[ht]
\caption{RAN Simulators Comparison}
\label{tab:RAN Simulators Comparison}
\centering
\begin{tabular}{p{0.23\columnwidth}p{0.30\columnwidth}p{0.19\columnwidth}}
\toprule
\textbf{Simulator} &  \textbf{Key Features} & \textbf{Limitations} \\
\midrule
srsRAN & LTE/5G & scalability,E2  \\
OAI & LTE/5G  &  scalability,E2 \\
my5G tester & 5G  & Non-O-RAN  \\
Simu5G & 5G analysis & Non-O-RAN  \\
5G-air & 5G Support & Non-O-RAN\\
MATLAB 5G & Simulink & Non-O-RAN  \\
NS2 & Dev Protocol  & Non-O-RAN  \\
NS3 & Active development & Non-O-RAN  \\
OMNeT++ & Modularity & Non-O-RAN  \\
OPNET & Detailed modeling & Non-O-RAN  \\
WISE & 5G Channel  & Non-O-RAN  \\
\bottomrule
\end{tabular}
\end{table}
\vspace{-1pt}

%% file: include/Handover_workflow.tex
\section{Handover in Next-G}
\begin{figure*}
\centering
\includegraphics[width=1\textwidth]{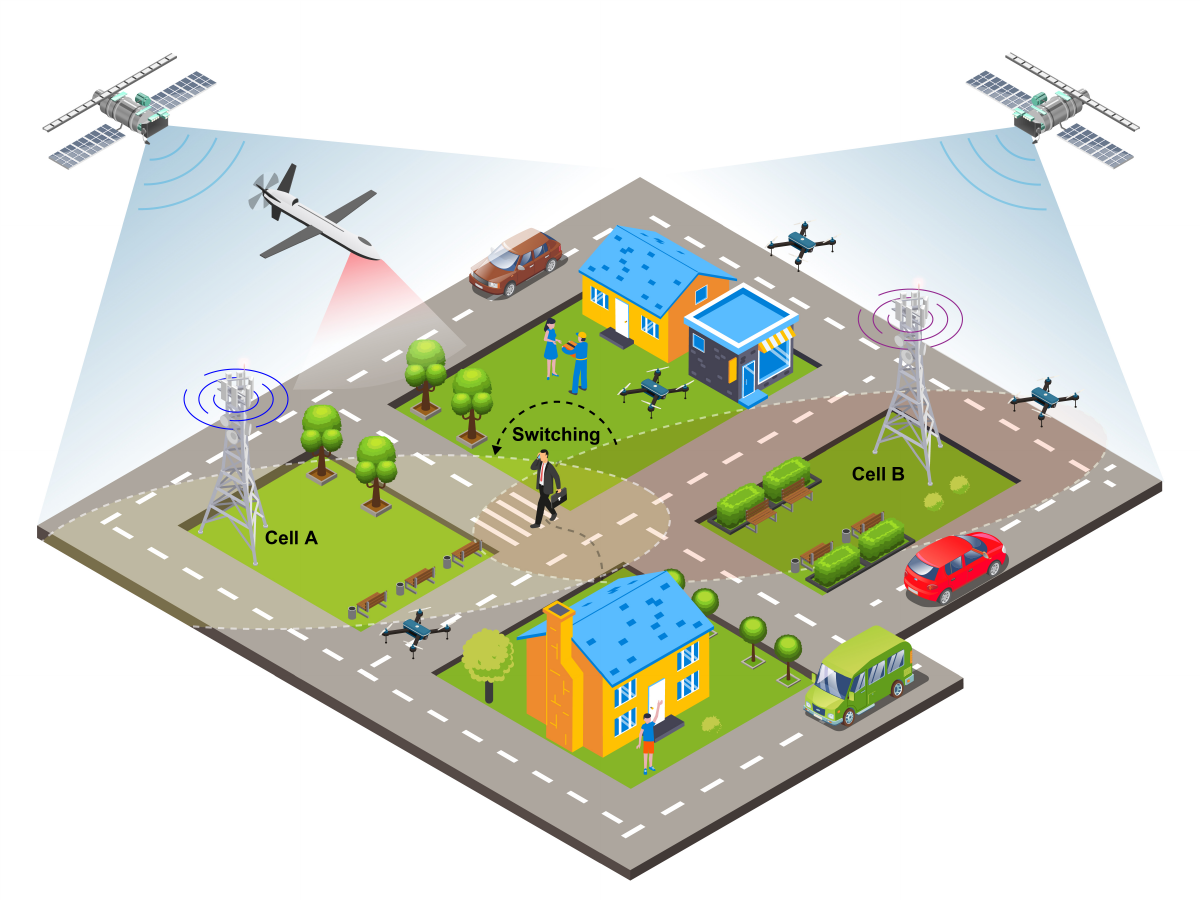}
\caption{Multiple Handover Scenarios in Future Networks}
\label{Handover-Next-G}
\vspace{-1pt}
\end{figure*}
Cellular communications require an efficient handover process to maintain service continuity. This process is crucial as users move between various network components, including sectors, cells, base stations, Centralized Units (CUs), and Distributed Units (DUs). As a bridge to 6G, 3GPP Release 19 focuses on refining mobility management and handover processes in commercial 5G networks. These processes can be triggered based on various scenarios, needs, and parameters, with the primary goal of minimizing service interruptions, maintaining service levels, and enhancing overall performance and revenue \cite{lin2023bridge}. Figure \ref{Handover-Next-G} illustrates multiple possible handover scenarios in 6G and next-generation networks. 

 In 6G networks, handover triggering events will become more complex to optimize network objectives such as load balancing, energy efficiency, and connection prioritization\cite{panitsas2024predictive}. Therefore, the handover decision-making process will depend on multiple parameters, requiring adaptable and intelligent algorithms.\newline
\vspace{-6pt}

\noindent Within the O-RAN architecture, handovers are classified into four main types: intra-gNB-DU, inter-gNB-DU, intra-gNB-CU, and inter-gNB-CU \cite{ORAN2021}. RANFusion adeptly supports these varied handover mechanisms, including soft handovers within the same cell and sector and hard handovers across different gNBs. This adaptability facilitates efficient transitions, including intra-gNB-DU handovers within the same gNB-DU and both inter-gNB-DU and intra-gNB-CU handovers across different gNB-DUs under the same gNB-CU. Moreover, inter-gNB-CU handovers, managed via the Xn or N2 interfaces \ref{Figure:ORAN-arc}, showcase the versatility of RANFusion within the disaggregated RAN structure of the O-RAN architecture, enhancing mobility management and resource optimization in next-generation cellular networks.

%% file: include/RANFusion_System_Design_and_Architecture.tex
\section{RANFusion: System Design and Architecture}
\begin{figure*}
\centering
\includegraphics[width=1\textwidth]{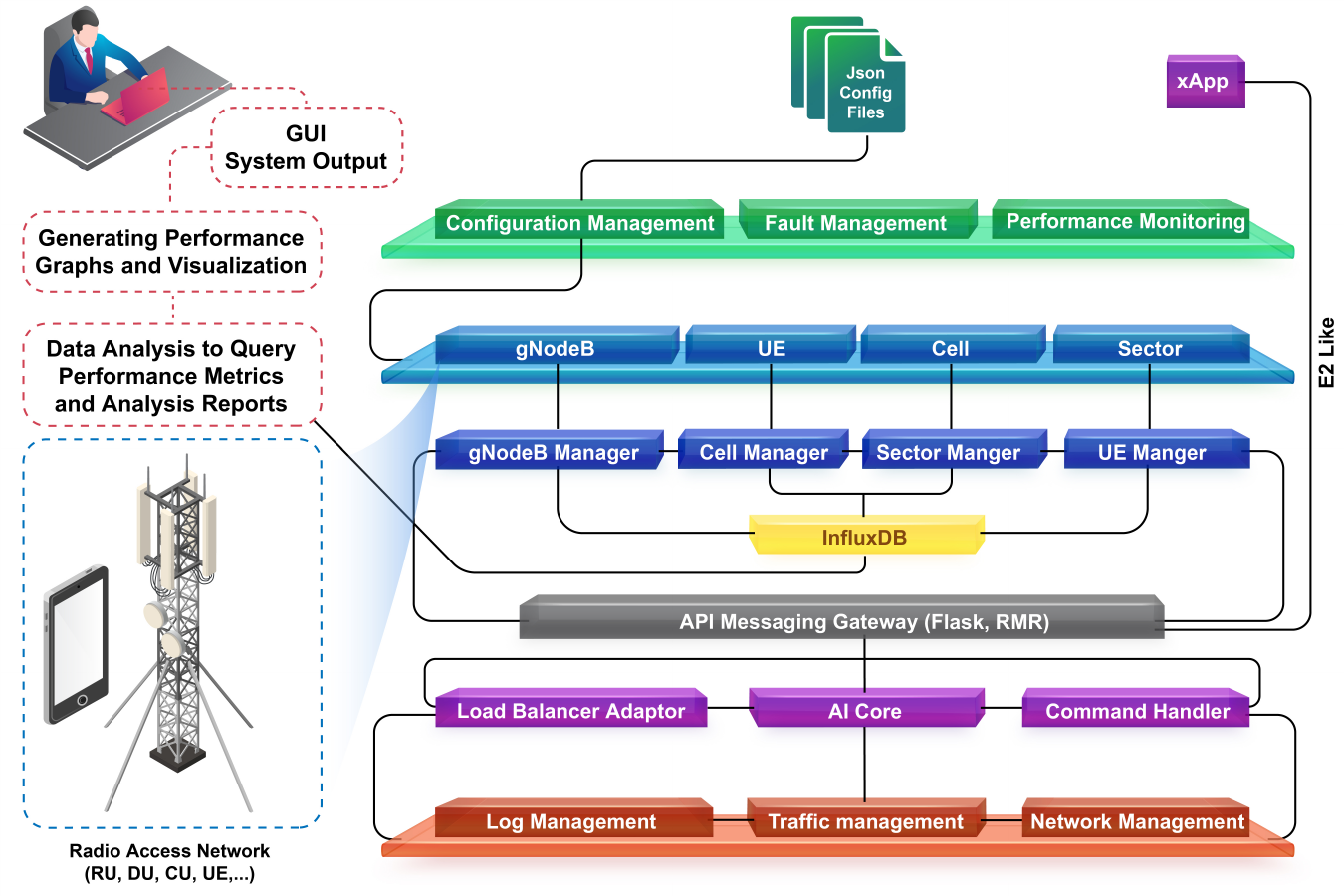}
\caption{Architecture of RANFusion}
\label{fig:ORAN-arc}
\vspace{-2pt}
\end{figure*}

\subsection{Requirement}
This simulator is compatible with Windows, MacOS, and Linux, requiring Python 3.x and InfluxDB 2.7.6 for its operation.
\subsection{Setup Networks}
Network setup requires specifying the number of cells, sectors, and gNodeBs in a JSON configuration. RANFusion facilitates customizable, scalable configurations from 1 to n sectors, cells, or gNodeBs. Each gNodeB is configured to support multiple cells, and each cell can encompass several sectors, with each sector capable of handling an unlimited number of User Equipment (UEs). Entity parameters are fully customizable via JSON. Figure \ref{fig:ORAN-arc} depicts the default radio architecture, mirroring commercial products, with 6 cells and 3 sectors per cell. Cell sectorization varies according to operator needs, environmental density, and network design considerations.
\subsection{Components}
The proposed architecture is detailed in Fig. \ref{fig:ORAN-arc}, with its major components described below:

\vspace{3pt}
\noindent \textbf{Configuration Management:}
RANFusion leverages four key JSON configuration files to define parameters for gNodeB, cell, sector, and UE configurations. This enables the customization of entities including cell id, sector id, bandwidth, latitude, and longitude, etc. Additionally, these files manage cell and sector capacities and define further parameters for UEs. The configuration management system reads and loads all configuration files, thereby initiating the 5G network RAN. In this setup, each gNodeB, Sector, cell, and UE is instantiated separately in memory and managed independently. 

\vspace{3pt}
\noindent \textbf{API Messaging Gateway:} RANFusion exposes a Flask API, which can enable users, external applications, xApps, or rApps to access and modify end-to-end entity parameters such as throughput, delay, and cell or sector capacity value during the simulation. This feature enables easier integration of RANFusion with O-RAN test beds or other necessary measurement tools.

\vspace{3pt}
\noindent \textbf{Traffic Management:} 
The traffic generator produces a variety of traffic types, including voice, video, gaming, IoT, and data, with customizable parameters such as packet size, bitrate, and interval. This facilitates the creation of conditions ranging from mild to extreme, aiding in resilience studies. The DatabaseManager component is responsible for storing real-time traffic metrics, which can be analyzed thoroughly.

\vspace{3pt}
\noindent \textbf{Network Management:} Network management consists of all entity nodes like sectors, cells, and gNodeBs, with all parameters for each entity defined in these components. Each of these nodes is controlled by its manager. UEs are dispatched across sectors using a "Round Robin with Fallback" strategy, which sequentially assigns UEs based on sector capacity and includes a fallback mechanism to manage resource constraints and specific demands. This strategy efficiently mitigates risks, streamlines decision-making, and prevents sector overloading.

\vspace{3pt}
\noindent \textbf{Database Management:} The database node manages all database actions, including recording all transactions between the UE and other nodes and defining and storing KPIs for each entity. Using InfluxDB, which is designed for time series data, it records metrics one second at a time in real-time.

\vspace{3pt}
\noindent \textbf{Loadbalancer:} 
The load balancer employs a proactive approach, where in-network loads are distributed among cells, sectors, and gNodeBs. By leveraging real-time metrics, this mechanism dynamically adjusts the distribution of UE connections. The effectiveness of this load-balancing strategy is assessed through several key metrics as defined below:

\begin{equation}
\scalebox{1.0}{
$\text{UE Count Load} = \left( \frac{\text{Number of Connected UEs}}{\text{Sector Capacity}} \right) \times 100$
}
\end{equation}

\begin{equation}
\scalebox{0.9}{
$\text{Throughput Load} = \left( \frac{\text{Total Capped Throughput}}{\text{Sector Maximum Throughput}} \right) \times 100$
}
\end{equation}

\begin{equation}
\scalebox{0.9}{
$\begin{aligned}
\text{Sector Load} = & (\text{COUNT\_WEIGHT} \times \text{UE Count Load}) \\
                     & + (\text{TP\_WEIGHT} \times \text{Throughput Load})
\end{aligned}$
}
\end{equation}

\begin{equation}
\text{Cell Load} = \frac{\sum_{i=1}^{n} \text{Sector Load}_i}{n}
\end{equation}

\begin{equation}
\text{gNodeB Load} = \frac{\sum_{i=1}^{m} \text{Cell Load}_i}{m}
\end{equation}

When the algorithm detects a potential risk of congestion, such as 80\% sector capacity, a handover process will be initiated. This process moves UEs to neighboring underloaded sectors, cells, or gNodeBs to mitigate the congestion risk. In case of any failure during this process, a rollback process is automatically initiated, returning UEs to its previous position.

\vspace{3pt}
\noindent \textbf{AI Core:} 
In the AI core, a variety of \textit{Handover Algorithms} and \textit{Mobility Models} are pooled to perform user-defined scenarios. The RANfusion load balancer can select and call several plug-and-play algorithms to precisely model mobile user behaviors and decision-making processes related to handover management. Performance evaluation and optimization are simplified with AI core modularity, and results are shown in the influxDB dashboard. Several studies, such as \cite{8812724}, \cite{8565842}, \cite{7994914}, and \cite{9195500}, include handover scenarios for various environments, which can be incorporated into the AI core pool.

\begin{figure*}[h]
\centering
\includegraphics[width=2\columnwidth]{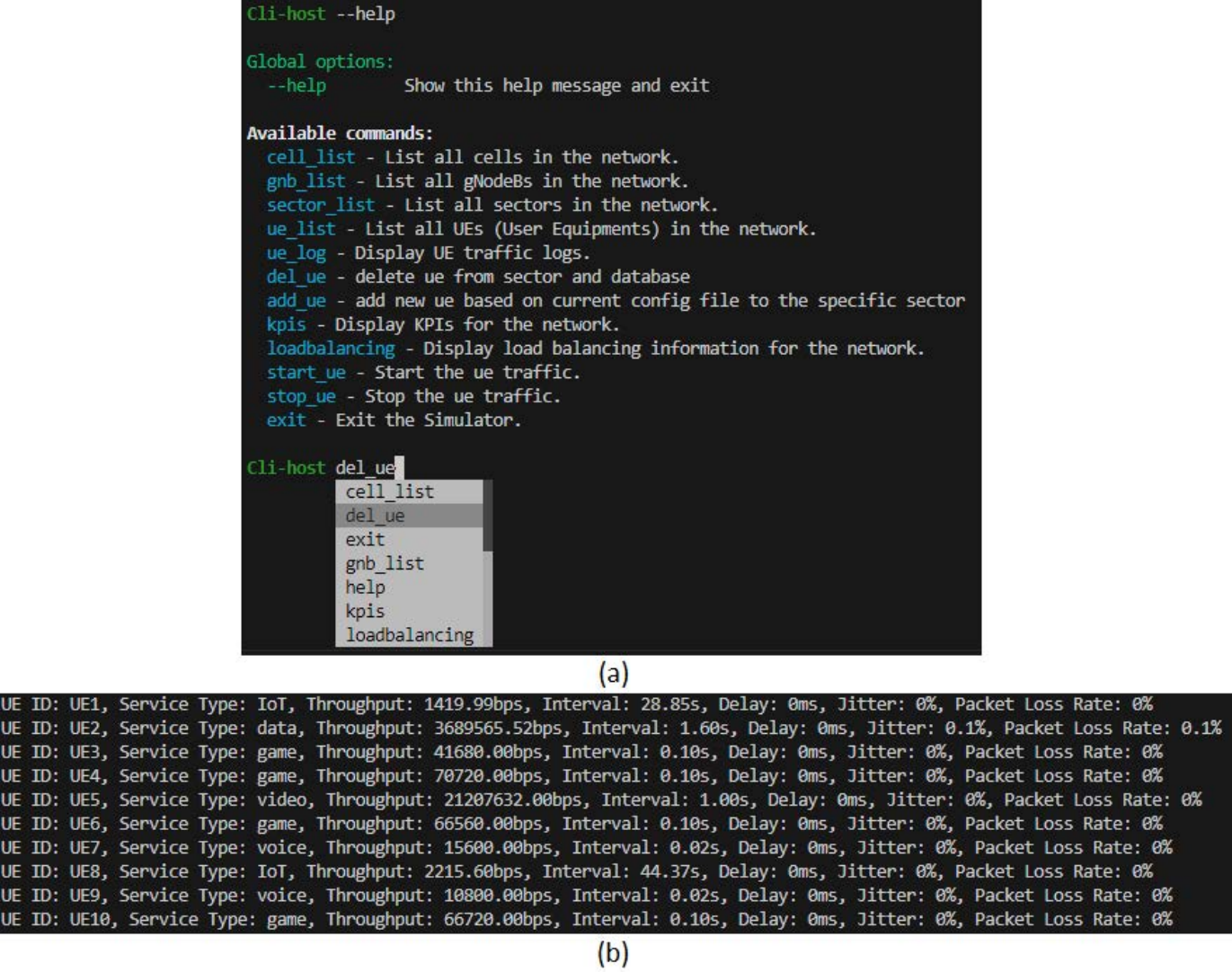}
\caption{The command-line user interface}
\label{figure:Command-line}
\end{figure*}

\vspace{3pt}
\noindent \textbf{Command Handler:}
The command handler node, which is shown in figure \ref{figure:Command-line} (a), is designed to act as a middleware for executing various UE activities (e.g., \texttt{add\_ue}, \texttt{del\_ue}), traffic control commands such as \texttt{start\_ue\_traffic} and \texttt{stop\_ue\_traffic}, and other related activities. When the simulator runs, this command line appears as a menu, and users can select and execute different commands, like \texttt{ue\_log} figure~\ref{figure:Command-line} (b), to see the results in the output or on the InfluxDB dashboard.

%% file: include/Experiment_Result.tex
\section{Experiment Results}
We designed an experiment to evaluate the performance of the RANFusion handover simulator in a realistic network environment with diverse user mobility patterns and traffic loads. The experimental setup consists of a network with three gNodeBs (gNBs) organized in a hexagonal layout. Each gNodeB is configured to support multiple cells, each encompassing several sectors, with sectors capable of handling an unlimited number of User Equipment. These sectors were capped at a throughput of 100 MB and accommodated User Equipment (UE) across various service classes, assigned randomly. The intersite distance between gNBs was established at 500 meters. 

\noindent Network management component uniformly placed 10 UEs across all cells (Sectors) and exhibited diverse usage patterns involving voice, data, video, gaming, and IoT services. The simulation was conducted over a span of 300 seconds to evaluate the sustained performance of the default handover protocol.

\vspace{2pt} \noindent To assess RANFusion’s load-balancing capabilities, we methodically increased the UE throughput in a specific sector (Sector A) via an API designed to simulate E2-like functionalities. This API supports a variety of actions, several of which are crucial for real-time adjustments during simulations, as outlined in the Command Handler section. During this test, the UE throughput load in Sector A was augmented by 10\% every 1 seconds until it reached 80\% of its sector capacity. This scenario simulates a rush-hour situation where a specific area experiences a surge in user traffic. Configured with a sector capacity threshold of 80\%, the RANFusion handover algorithm initiates the handover process when the load within a sector surpasses this threshold. As a part of this process, selected UEs are transitioned either to an adjacent sector or to a neighboring cell that has sufficient capacity. The handover decision considers the available neighboring sectors and cells, as well as the load conditions of the candidate cells. The API used to change UE throughput is also available to any xApp or rApp. Other parameters, such as those for UE, cell, sector, or gNodeB, including UE signal strength, can be defined or modified in the handover algorithm or the AI core. This makes it easy to change certain parameters that have to do with gNodeBs, cells, sectors, UEs, or traffic flows. This capability highlights the flexibility and adaptability of our system in dynamic network conditions. We evaluated the handover algorithm's performance based on several key metrics:

\vspace{3pt}
\noindent\textit{Handover Success Rate (HSR):} 
Handover success rate versus attempted handovers(ratio).

\vspace{3pt}
\noindent\textit{Handover Failure Rate (HFR):} Failure rate of handovers (ratio).

\vspace{3pt}
\noindent\textit{Average UE Throughput:} The average data rate experienced by the UEs in the network.

\vspace{3pt}
\noindent\textit{Cell and Sector Load Distribution:} The distribution of traffic load across all cells in the network.

\begin{figure}[h]
\centering
\includegraphics[width=\columnwidth]{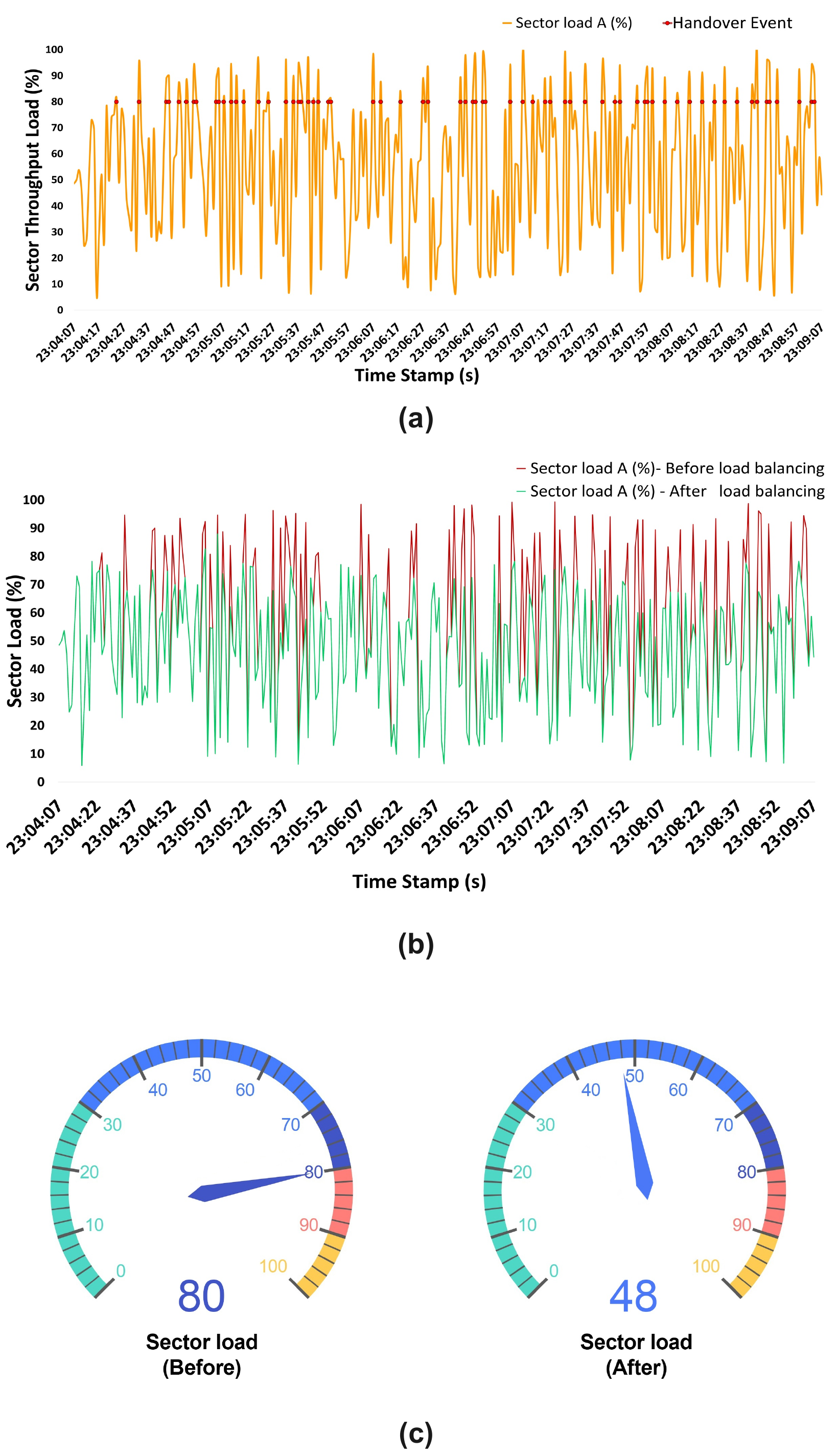}
\caption{(a) Handover peak during the time, (b) load ratio comparison before and after load balancing, showing traffic load on sector A before and after load balancing (c).}
\label{figure:handover_result}
\end{figure}

\noindent The results demonstrate the effectiveness of the RANFusion handover algorithm in managing the network load and ensuring a high handover success rate. The algorithm successfully offloads UEs from the congested target sector/cell to the neighboring sector/cell (soft handover), resulting in a more balanced load distribution across the network. The average sector load is maintained at 48\%, with the maximum cell load not exceeding 80\%, indicating efficient resource utilization.

Figure \ref{figure:handover_result} visualizes the sector load distribution after applying the RANFusion default handover algorithm. Graph (a) clearly shows the handover peak and illustrates how the algorithm redistributes the UE load from the congested sector to other sectors or cells, achieving a more uniform load distribution. The algorithm effectively manages diverse mobility patterns and varying traffic loads, ensuring seamless connectivity for the UEs. Graph (b) also shows the before and after load balancing of sector A, and graph (c) presents the Grafana output schematic, which is connected to InfluxDB to monitor the sector load during the handover event. Other supported parameters and sample KPIs are also presented in Table \ref{table:ranfusion_parameters}.

\begin{table}[ht]
\caption{RANFusion Simulation Sample Parameters }
\label{table:ranfusion_parameters}
\centering
\begin{tabular}{ll}
\toprule
\textbf{Parameter} & \textbf{Value} \\
\midrule
Handover Count & 98    (int) \\
Handover Latency & 0.5  (s) \\
Calculate Handover Failure Rate (HFR) & 0.1 (\%) \\
Handover Success Rate (HSR) & 99 (\%) \\
UE Throughput & 40.3 (MBps) \\
Network Delay & 0.1 (s) \\
UE Jitter & 0.1 (ms)\\
UE Packet loss & 0.2 (\%)\\
UE Delay & 0.01 (s)\\
Cell Load& 53 (\%) \\
Sector Load& 65  (\%)  \\
Network Load& 35.8 (\%)  \\
\bottomrule
\end{tabular}
\end{table}


%% file: include/Future_research_directions_and_potential_enhancements_to_RANFusion.tex
\section{Future research directions}
This paper addresses the need for an effective RAN and O-RAN handover simulator. Utilizing an InfluxDB graph to store and visualize results, which initially focuses on the crucial issue of sector or cell congestion, Future enhancements will address the current limitations by establishing a direct connection between the simulator and the RIC through the E2 interface. We aim to integrate and dockerize the system with OAIC-T \cite{tang2023ai} to develop a comprehensive O-RAN testbed framework.

%% file: main.bbl
\begin{thebibliography}{10}
\providecommand{\url}[1]{#1}
\csname url@samestyle\endcsname
\providecommand{\newblock}{\relax}
\providecommand{\bibinfo}[2]{#2}
\providecommand{\BIBentrySTDinterwordspacing}{\spaceskip=0pt\relax}
\providecommand{\BIBentryALTinterwordstretchfactor}{4}
\providecommand{\BIBentryALTinterwordspacing}{\spaceskip=\fontdimen2\font plus
\BIBentryALTinterwordstretchfactor\fontdimen3\font minus \fontdimen4\font\relax}
\providecommand{\BIBforeignlanguage}[2]{{%
\expandafter\ifx\csname l@#1\endcsname\relax
\typeout{** WARNING: IEEEtran.bst: No hyphenation pattern has been}%
\typeout{** loaded for the language `#1'. Using the pattern for}%
\typeout{** the default language instead.}%
\else
\language=\csname l@#1\endcsname
\fi
#2}}
\providecommand{\BIBdecl}{\relax}
\BIBdecl

\bibitem{lin2023fundamentals}
\BIBentryALTinterwordspacing
Lin,~X., Zhang,~J., Liu,~Y., and Kim,~J., \emph{Fundamentals of 6G Communications and Networking}, ser. Signals and Communication Technology.\hskip 1em plus 0.5em minus 0.4em\relax Springer International Publishing, 2023. [Online]. Available: \url{https://books.google.com/books?id=a-HoEAAAQBAJ}
\BIBentrySTDinterwordspacing

\bibitem{10279738}
Mohammadi,~H., Marojevic,~V., and Shang,~B., ``Analysis of reinforcement learning schemes for trajectory optimization of an aerial radio unit,'' in \emph{ICC 2023 - IEEE International Conference on Communications}, 2023, pp. 6423--6428.

\bibitem{panitsas2024predictive}
Panitsas,~I., Mudvari,~A., Maatouk,~A., and Tassiulas,~L., ``Predictive handover strategy in 6g and beyond: A deep and transfer learning approach,'' 2024.

\bibitem{DAO2024110370}
\BIBentryALTinterwordspacing
Dao,~N.-N., Tu,~N.~H., Hoang,~T.-D., Nguyen,~T.-H., Nguyen,~L.~V., Lee,~K., Park,~L., Na,~W., and Cho,~S., ``A review on new technologies in 3gpp standards for 5g access and beyond,'' \emph{Computer Networks}, p. 110370, 2024. [Online]. Available: \url{https://www.sciencedirect.com/science/article/pii/S1389128624002020}
\BIBentrySTDinterwordspacing

\bibitem{Haghrah2023}
\BIBentryALTinterwordspacing
Haghrah,~A., Abdollahi,~M.~P., and Azarhava,~H., ``A survey on the handover management in 5g-nr cellular networks: aspects, approaches and challenges,'' \emph{Journal of Wireless Communications and Networking}, vol. 2023, no.~52, 2023. [Online]. Available: \url{https://doi.org/10.1186/s13638-023-02261-4}
\BIBentrySTDinterwordspacing

\bibitem{PATRICIELLO2019101933}
\BIBentryALTinterwordspacing
Patriciello,~N., Lagen,~S., Bojovic,~B., and Giupponi,~L., ``An e2e simulator for 5g nr networks,'' \emph{Simulation Modelling Practice and Theory}, vol.~96, p. 101933, 2019. [Online]. Available: \url{https://www.sciencedirect.com/science/article/pii/S1569190X19300589}
\BIBentrySTDinterwordspacing

\bibitem{9211504}
Nardini,~G., Sabella,~D., Stea,~G., Thakkar,~P., and Virdis,~A., ``Simu5g–an omnet++ library for end-to-end performance evaluation of 5g networks,'' \emph{IEEE Access}, vol.~8, pp. 181\,176--181\,191, 2020.

\bibitem{7801585}
Dominguez-Bolano,~T., Rodriguez-Pineiro,~J., Garcia-Naya,~J.~A., and Castedo,~L., ``The gtec 5g link-level simulator,'' in \emph{2016 1st International Workshop on Link- and System Level Simulations (IWSLS)}, 2016, pp. 1--6.

\bibitem{SILVEIRA2022109301}
\BIBentryALTinterwordspacing
Silveira,~L.~B., {de Resende},~H.~C., Both,~C.~B., Marquez-Barja,~J.~M., Silvestre,~B., and Cardoso,~K.~V., ``Tutorial on communication between access networks and the 5g core,'' \emph{Computer Networks}, vol. 216, p. 109301, 2022. [Online]. Available: \url{https://www.sciencedirect.com/science/article/pii/S1389128622003528}
\BIBentrySTDinterwordspacing

\bibitem{MARTIRADONNA2020107314}
\BIBentryALTinterwordspacing
Martiradonna,~S., Grassi,~A., Piro,~G., and Boggia,~G., ``Understanding the 5g-air-simulator: A tutorial on design criteria, technical components, and reference use cases,'' \emph{Computer Networks}, vol. 177, p. 107314, 2020. [Online]. Available: \url{https://www.sciencedirect.com/science/article/pii/S1389128619317347}
\BIBentrySTDinterwordspacing

\bibitem{boeira2024calibrated}
Boeira,~C., Hasan,~A., Papry,~K., Ju,~Y., Zhu,~Z., and Haque,~I., ``A calibrated and automated simulator for innovations in 5g,'' 2024.

\bibitem{lin2023bridge}
Lin,~X., ``The bridge toward 6g: 5g-advanced evolution in 3gpp release 19,'' 2023.

\bibitem{ORAN2021}
{O-RAN Working Group 1}, ``O-ran architecture description v11.00,'' O-RAN Alliance, {Technical Specification} ORAN.WG1.O-RAN-Architecture-Description-v11.00, 2024.

\bibitem{8812724}
Tayyab,~M., Gelabert,~X., and Jäntti,~R., ``A survey on handover management: From lte to nr,'' \emph{IEEE Access}, vol.~7, pp. 118\,907--118\,930, 2019.

\bibitem{8565842}
Jain,~A., Lopez-Aguilera,~E., and Demirkol,~I., ``Evolutionary 4g/5g network architecture assisted efficient handover signaling,'' \emph{IEEE Access}, vol.~7, pp. 256--283, 2019.

\bibitem{7994914}
Bilen,~T., Canberk,~B., and Chowdhury,~K.~R., ``Handover management in software-defined ultra-dense 5g networks,'' \emph{IEEE Network}, vol.~31, no.~4, pp. 49--55, 2017.

\bibitem{9195500}
Shayea,~I., Ergen,~M., Hadri~Azmi,~M., Aldirmaz~Çolak,~S., Nordin,~R., and Daradkeh,~Y.~I., ``Key challenges, drivers and solutions for mobility management in 5g networks: A survey,'' \emph{IEEE Access}, vol.~8, pp. 172\,534--172\,552, 2020.

\bibitem{tang2023ai}
Tang,~B., Shah,~V.~K., Marojevic,~V., and Reed,~J.~H., ``Ai testing framework for next-g o-ran networks: Requirements, design, and research opportunities,'' \emph{IEEE Wireless Communications}, vol.~30, no.~1, pp. 70--77, 2023.

\end{thebibliography}
